\documentclass[useAMS,usenatbib]{mn2e}
\usepackage{natbib}
\usepackage{amsmath,amsfonts}
\usepackage{epsfig}
\usepackage[table]{xcolor}
\usepackage{mathptmx}
\usepackage{longtable}

\def\reff@jnl#1{{\rm#1\/}}

\def\aj{\reff@jnl{AJ}}                  
\def\araa{\reff@jnl{ARA\&A}}            
\def\apj{\reff@jnl{ApJ}}                        
\def\apjl{\reff@jnl{ApJ}}               
\def\apjs{\reff@jnl{ApJS}}              
\def\ao{\reff@jnl{Appl.Optics}}         
\def\apss{\reff@jnl{Ap\&SS}}            
\def\aap{\reff@jnl{A\&A}}                       
\def\apjl{\reff@jnl{ApJ}}               
\def\aapr{\reff@jnl{A\&A~Rev.}}         
\def\aaps{\reff@jnl{A\&AS}}             
\def\azh{\reff@jnl{AZh}}                        
\def\baas{\reff@jnl{BAAS}}              
\def\jrasc{\reff@jnl{JRASC}}            
\def\memras{\reff@jnl{MmRAS}}           
\def\mnras{\reff@jnl{MNRAS}}            
\def\pra{\reff@jnl{Phys. Rev. A}}         
\def\prb{\reff@jnl{Phys. Rev. B}}         
\def\prc{\reff@jnl{Phys. Rev. C}}         
\def\prd{\reff@jnl{Phys. Rev. D}}         
\def\prl{\reff@jnl{Phys. Rev. Lett}}      
\def\pasp{\reff@jnl{PASP}}              
\def\pasj{\reff@jnl{PASJ}}              
\def\qjras{\reff@jnl{QJRAS}}            
\def\skytel{\reff@jnl{S\&T}}            
\def\solphys{\reff@jnl{Solar~Phys.}}    
\def\sovast{\reff@jnl{Soviet~Ast.}}     
\def\ssr{\reff@jnl{Space~Sci.Rev.}}     
\def\zap{\reff@jnl{ZAp}}                        
\def\nat{\reff@jnl{Nature}}             

\def\p#1by#2{{\partial{#1} \over \partial{#2}}}
\def\pp#1by#2#3{{\partial^2{#1} \over \partial{#2}\partial{#3}}}
\def\d#1by#2{{{\rm d}{#1} \over {\rm d}{#2}}}
\def\dd#1by#2#3{{{\rm d}^2{#1} \over {\rm d}{#2}{\rm d}{#3}}}

\title[Low frequency broadband flux scale]{A broadband flux scale for low frequency radio telescopes}

\label{firstpage}

\author[Scaife \& Heald]{
 Anna M. M. Scaife$^{1}$\thanks{email: a.scaife@soton.ac.uk} \&
 George H. Heald$^{2}$.
 \vspace{0.03in}\\
$^1$ School of Physics \& Astronomy, University of Southampton, Highfield, Southampton, SO17 1BJ\\
$^2$ ASTRON, the Netherlands Institute for Radio Astronomy, Postbus 2, 7990 AA, Dwingeloo, The Netherlands
}

\date{Accepted ---; received ---; in original form \today}

\pagerange{\pageref{firstpage}--\pageref{lastpage}}

\pubyear{2010}

\begin{document}
\maketitle

\begin{abstract}
We present parameterized broadband spectral models valid at frequencies between 30-300\,MHz for six bright radio sources selected from the 3C survey, spread in Right Ascension from $0-24$\,hours. For each source, data from the literature are compiled and tied to a common flux density scale. These data are then used to parameterize an analytic polynomial spectral calibration model. The optimal polynomial order in each case is determined using the ratio of the Bayesian evidence for the candidate models. Maximum likelihood parameter values for each model are presented, with associated errors, and the percentage error in each model as a function of frequency is derived. These spectral models are intended as an initial reference for science from the new generation of low frequency telescopes now coming on line, with particular emphasis on the Low Frequency Array (LOFAR).
\end{abstract}


\section{Introduction}

In order to quantitatively combine and contrast data from independent telescopes and surveys, often at multiple frequencies, it is necessary to have a standard calibration scale to form comparisons. This is especially important at frequencies below 300\,MHz and above 15\,GHz where the widely used Baars et~al. (1977) radio flux density scale is incomplete. For the new generation of low frequency telescopes such as the Low Frequency Array (LOFAR; van~Haarlem et~al. in prep) it is becoming increasingly necessary to provide a broadband spectral reference for initial science, so that both archival and future measurements can be quantitatively compared to these new data. In addition to an absolute scaling, such telescopes require a well-defined set of calibrators spread in right ascension (RA) to allow for quasi-simultaneous broadband calibration of field observations. Here we present a set of parameterized models for six broadband calibrators covering frequencies from $30-300$\,MHz and RAs from $0-24$\,hrs. We focus on the northern sky, and in particular on the applicability to LOFAR. This set of calibrators forms a flux scale that will be the basis of a major effort to develop an all-sky, broadband calibration catalog. The initial description given here will be continuously refined as new LOFAR data accumulate.

\section{Calibration of low frequency telescopes}

Radio interferometers operating at low frequencies face a substantial calibration challenge. Strong ionospheric phase corruptions are common, especially below 100~MHz. For large-scale survey work in particular, it is important that the processing of raw visibility data from the telescope can be automated. In order to jump-start such an automatic calibration and imaging process for any arbitrary field, a pre-existing model of the brightest sources in the field of view is required. Such a model must be intrinsically frequency dependent, since modern radio telescopes are inherently broadband in nature, with tremendous fractional bandwidths. For example, LOFAR routinely observes from $30-240$~MHz, and is capable of observing as low as 10~MHz. Over such a broad range, the flux scales must be tied to a well-understood set of reference sources with spectral energy distributions that are well understood across the full bandpass. In the case of LOFAR, the production of such an all-sky broadband catalog is the key goal of the Multifrequency Snapshot Sky Survey (MSSS; Heald et al. in prep).

The reference sources which form the basis of the broadband flux scale must be selected for suitability as high-quality calibration targets. Several factors are relevant. First, the source should dominate the visibility function. In addition to high flux density, separation of contaminating flux from sources away from the pointing centre (`off-beam') can be improved in two further ways (i) averaging in time and frequency to smear out the contributions of off-beam sources on longer baselines; and (ii) the ``demixing'' technique (van der Tol et al. 2007), which has been adopted for use with LOFAR data. Secondly, the source should be compact compared to the angular resolution of the instrument, to allow simple morphological calibration models. Well-known sources such as Cyg A and Cas A have extremely complex morphologies, making calibration of an array with arcsecond angular resolution difficult. Thirdly, these calibrators must be spread in right ascension
(RA) to allow for quasi-simultaneous broadband calibration with
field observations.

With these considerations in mind we searched the 3C (Edge et~al. 1959) and revised 3C (3CR; Bennet et~al. 1962) catalogues for an initial list of bright compact sources, with the criteria that (1) they must be at declinations greater than 30$^{\circ}$, (2) they must have a flux density at 178\,MHz greater than 20\,Jy and (3) they must have an angular diameter less than 20\,arcseconds (compact compared to the naturally weighted resolution of the Dutch LOFAR array). 
\begin{figure*}
\centerline{\includegraphics[width=0.27\textwidth]{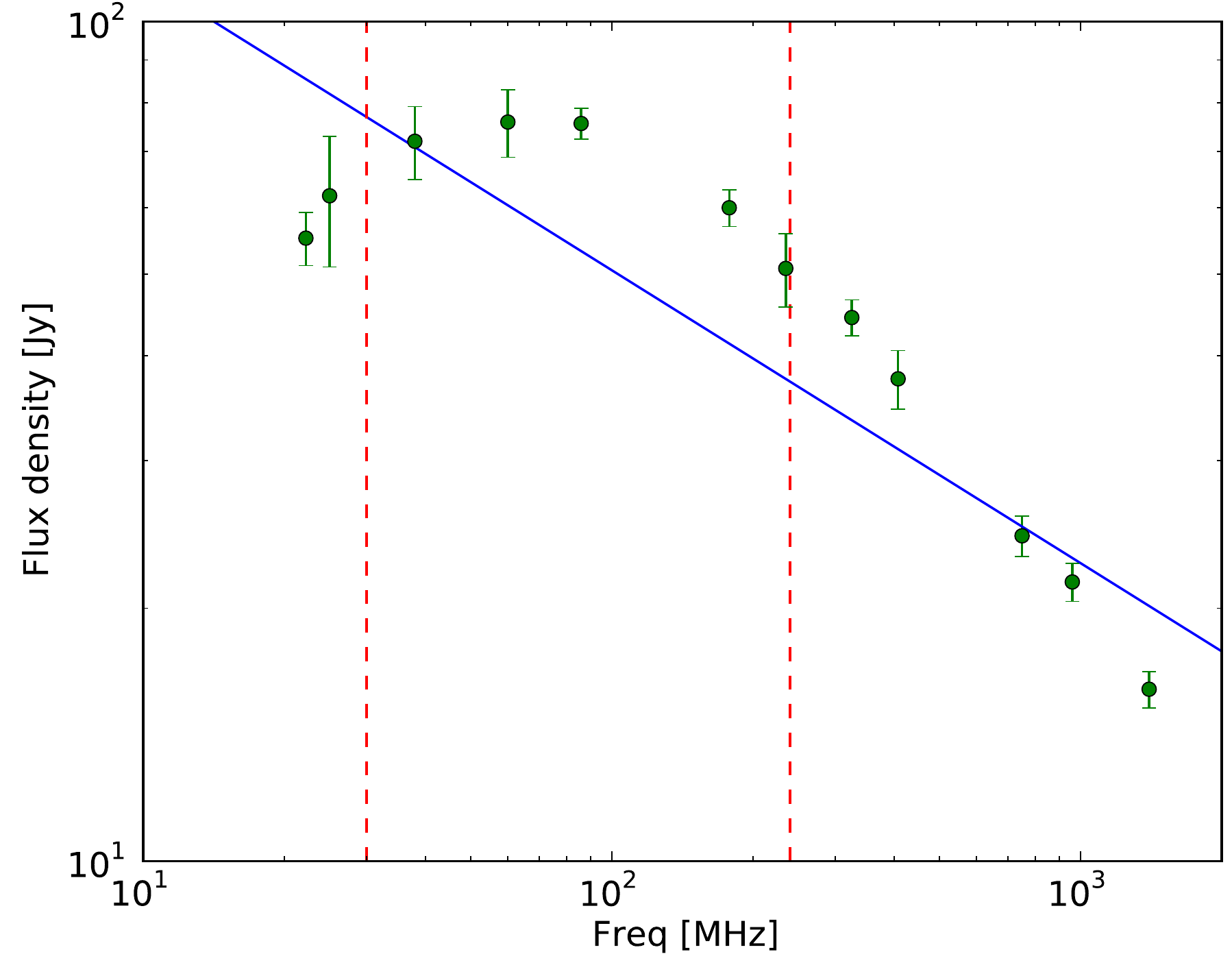} \includegraphics[width=0.25\textwidth]{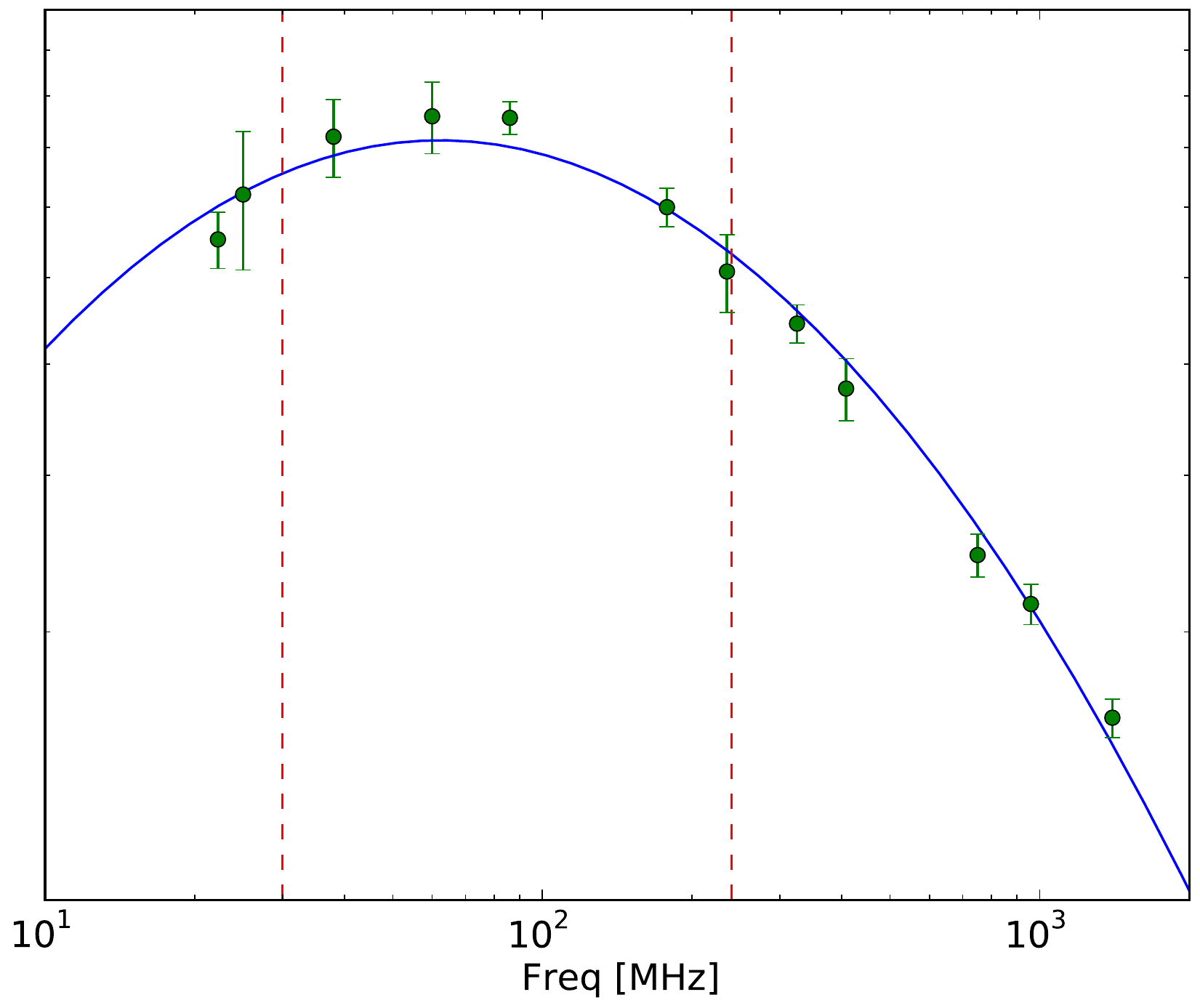} \includegraphics[width=0.25\textwidth]{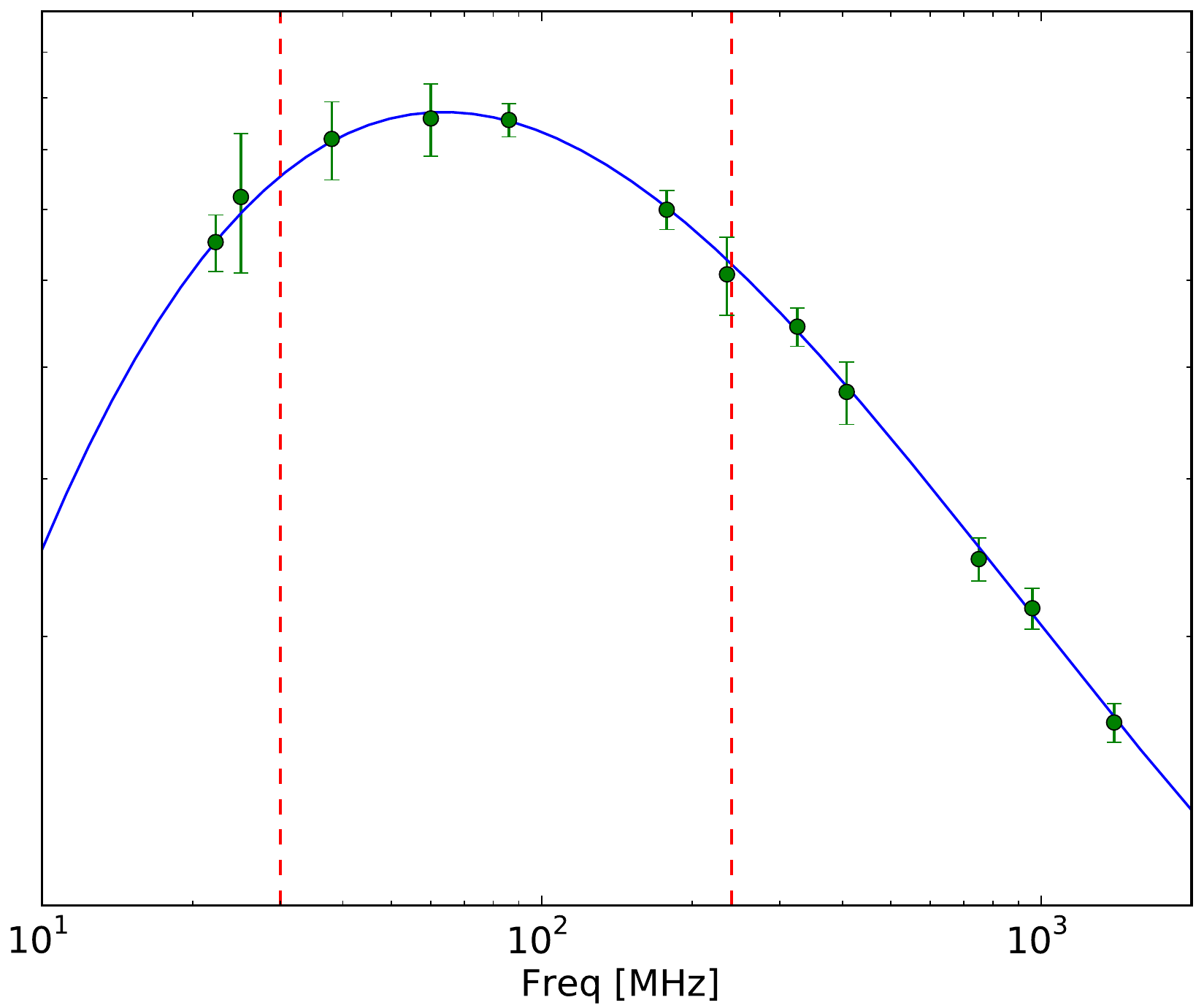} \includegraphics[width=0.25\textwidth]{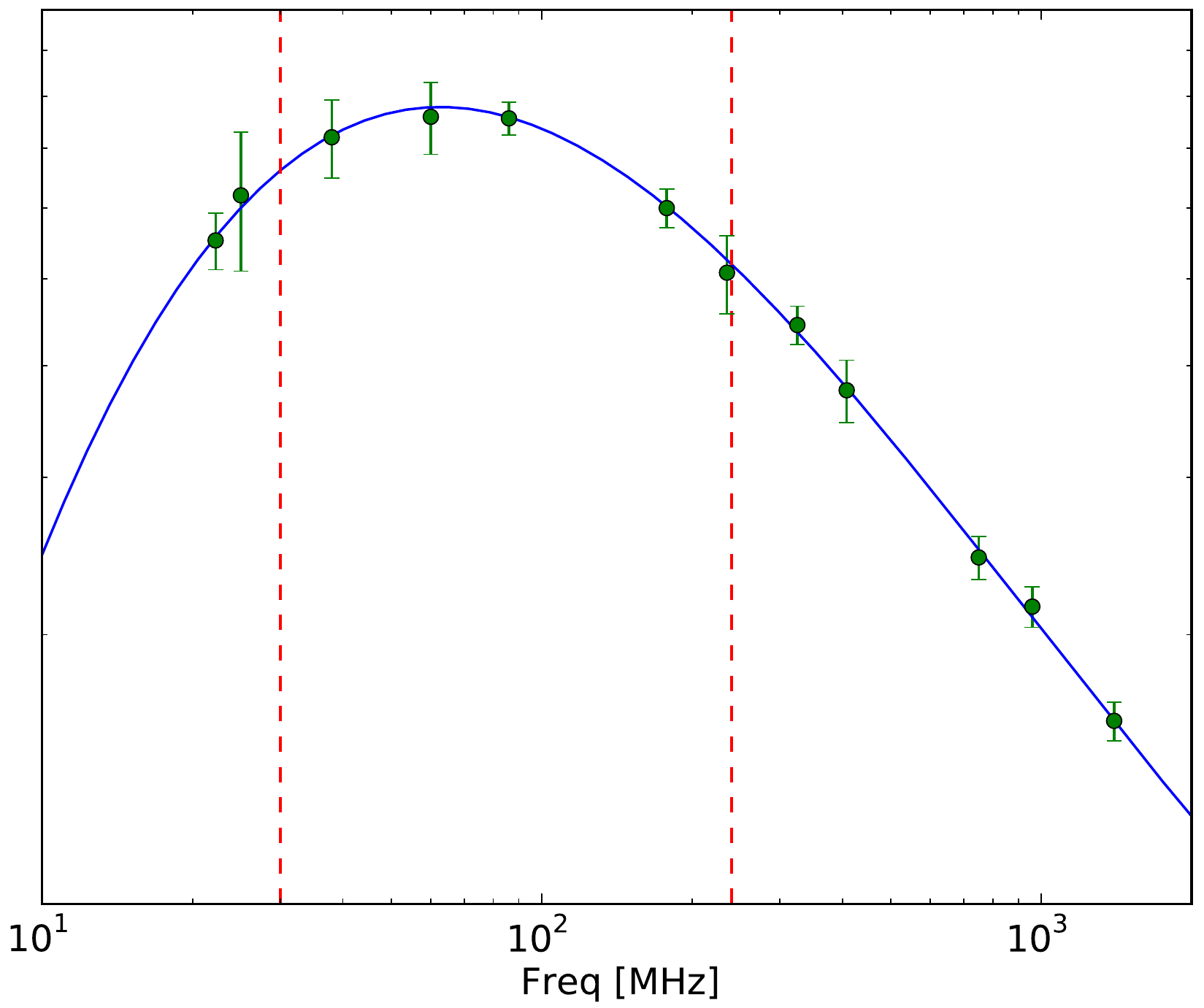}}

\centerline{ (a) \hskip 0.25\textwidth (b) \hskip 0.25\textwidth (c) \hskip 0.25\textwidth (d) }
\caption{3C48 SED. (a) Data fitted with linear ($1^{\circ}$) model, (b) data fitted with second order model, (c) data fitted with third order model and (d) data fitted with fourth order model. ML parameters for each fit are listed in Table~\ref{tab:res}. Dashed lines indicate the upper and lower bounds of the LOFAR frequency band. \label{fig:3c48}}
\end{figure*}

These criteria result in an initial sample of six sources, of which we exclude one based on other data from the literature showing more substantial extension than indicated in 3CR (3C69; Pooley \& Henbest 1974) and we include one additional source based on other data from the literature indicating that the extension listed in 3CR is an overestimate (3C286; e.g. Pearson et~al. 1985). The final sample is listed in Table~\ref{tab:cals}. Source extensions from 3CR are listed in Column [5] for each object. We note that high resolution observations (e.g. Akujar \& Garrington 1995) confirm that the source structure in 3C48 is on sub-arcsecond scales, whilst 3C147, 3C286 and 3C295 have structure on scales $<5''$. 3C196 has two dominant components separated by about $6''$, as well as complex diffuse structure with a (precessing) jet morphology (Reid et~al. 1995). The structure in 3C380 is known to be dominant on scales of $\approx 16''$, making it the most extended object in this sample (Reid et~al. 1995). 

\section{Flux Scales}

The data used for spectrum fitting are listed in Table~\ref{tab:refs}. In order to provide a common flux scaling, these data have been revised onto the flux scale of Roger, Bridle \& Costain (1973; hereafter RBC) below 325\,MHz. This scale has been chosen to avoid the suggested issues (e.g. Rees 1990a) with the secular decrease in the flux density of Cas~A at low frequencies ($<100$\,MHz) inherent in the widely used Baars et~al. (1977; B77) scale.

At low radio frequencies most data are tied to the RBC or Kellerman, Pauliny-Toth \& Williams (1969; KPW) scales. The correction factors for moving between these scales at $\nu<325$\,MHz are listed in Table.~\ref{tab:refs}. At $\nu>325$\,MHz the RBC and KPW scales are in agreement and consequently such data, where calibrated on the B77 scale, are corrected using a polynomial fit to the correction factors listed in B77 onto the KPW scale. Data from WENSS (Rengelink et~al. 1997) have been corrected using an average correction factor to bring them onto the B77 scale and a further scaling to bring them onto the RBC scale. The 6C, 8C and MIYUN surveys are calibrated on the RBC scale in their original form, and the Bologna survey (Colla et~al. 1970) is calibrated on the KPW scale which is consistent with the RBC at 408\,MHz. Data from Aslanyan et~al. (1968) are scaled using the ratio of the stated flux densities for the calibrator sources (3C348 \& 3C353) in the original paper to the predicted values at 60\,MHz from the spectral models for these sources in RBC. Data from Scott \& Shakeshaft (1971) are corrected onto the scale of Artyukh et~al. (1969) and then onto the RBC scale using the factors listed in Tables~III~\&~IV of RBC, this is subject to the caveat that the difference in flux densities from 81.5 to 86\,MHz is assumed to be negligible compared to the uncertainty in these factors ($\approx 3$\,per~cent). Where applied, the scaling factors in each case are listed in Table~\ref{tab:refs}. The original flux densities for the sources from the 3C catalogue (Edge et~al. 1959) have not been included in the model fitting. The large size of the errors associated to these data is such that they have no influence on the parameter estimation. 

Additional data are available at $12.6 - 25$\,MHz from the UTR-1 telescope (Braude et~al. 1970a;b), calibrated on the Gravoko scale. These data have not been used in the fitting, primarily because the discrepancy between the Gravoko and RBC flux scales is not only frequency but also flux density dependent and there is no complete revision scale available. For a discussion see RBC.

\begin{table}
\begin{center}
\caption{Calibration Source Sample. \label{tab:cals}}
\begin{tabular}{lcccc}
\hline\hline
Source & RA & Dec & $S_{3C}^{\dagger}$ & $\Delta\theta^{\ast}$ \\
       & (J2000) & (J2000) & (Jy) & (arcsec)\\
\hline
3C48 & 01 37 41.3 & +33 09 35 & $50\pm11$ & $<1$ \\
3C147& 05 42 36.1 & +49 51 07 & $63\pm12$ & $<12$ \\
3C196& 08 13 36.0 & +48 13 03 & $66\pm20$ & $<12$ \\
3C286& 13 31 08.3 & +30 30 33 & $21^{\ast}$ & $<3^{\ddagger}$ \\
3C295& 14 11 20.5 & +52 12 10 & $74\pm15$ & $<12$ \\
3C380& 18 29 31.8 & +48 44 46 & $70\pm10$ & $<20$ \\
\hline
\end{tabular}
\begin{minipage}{0.5\textwidth}
$^{\ast}$ values from the revised 3C catalogue (3CR; Bennet 1962).

$^{\dagger}$ unadjusted flux densities at 159\,MHz \hskip .05in $^{\ddagger}$ Pearson et~al. (1985).
\end{minipage}
\end{center}
\end{table}

\begin{table}
\begin{center}
\caption{References for data used in spectral fitting. Column [1] frequency; column [2] reference; column [3] correction factor applied to original data for conversion to RBC flux scale. \label{tab:refs}}
\begin{tabular}{llc}
\hline\hline
Freq. & Ref. & factor \\
\hline
10\,MHz & Bridle \& Purton 1968 & 1.20$^{\ast}$ \\
        & Roger, Bridle \& Costain 1973 & - \\
22.25\,MHz & Roger, Costain \& Lacey 1969 & 1.15$^{\ast}$\\
           & Roger, Bridle \& Costain 1973 & - \\
38\,MHz & Kellerman, Pauliny-Toth \& Williams 1969 & 1.18$^{\ast}$ \\
        & Rees 1990 (8C) & - \\ 
60\,MHz & Aslanyan et~al. 1968 & 1.04$^{\ddagger}$ \\
81.5\,MHz & Scott \& Shakeshaft 1971 & 0.90$^{\ddagger}$ \\
86\,MHz & Artyukh et~al. 1969 & 0.94$^{\ast}$ \\
151\,MHz & Baldwin et~al. 1985 (6C)  & - \\
178\,MHz & Kellerman, Pauliny-Toth \& Williams 1969 & 1.09$^{\ast}$ \\
232\,MHz & Zhang et~al. 1997 (MIYUN) & - \\
325\,MHz & Rengelink et~al. 1997 (WENSS) & 0.90$^{\ddagger}$ \\
408\,MHz & Colla et~al. 1970 & - \\
750\,MHz & Kellerman, Pauliny-Toth \& Williams 1969 & - \\
960\,MHz & Kovalev et~al. 1997 & 0.96$^{\dagger}$ \\
1400\,MHz & Kellerman, Pauliny-Toth \& Williams 1969 & - \\
\hline
\end{tabular}
\begin{minipage}{0.5\textwidth}
$^{\ast}$ from RBC; \hskip .1in $^{\dagger}$ from B77; \hskip .1in $^{\ddagger}$ see text for details.
\end{minipage}
\end{center}
\end{table}

\section{Spectral Model}

A spectral model of the form
\begin{equation}
\nonumber \log S = \log A_0 + A_1 \log \nu + A_2 \log^2 \nu + \cdots
\end{equation}
was used. The model was applied in linear frequency space, i.e.
\begin{equation}
\nonumber S [{\rm Jy}] = A_0\prod_{i=1}^{N}{ 10^{A_i \log^i [\nu/150\,{\rm MHz}]} },
\end{equation}
\begin{figure*}
\centerline{\includegraphics[width=0.33\textwidth]{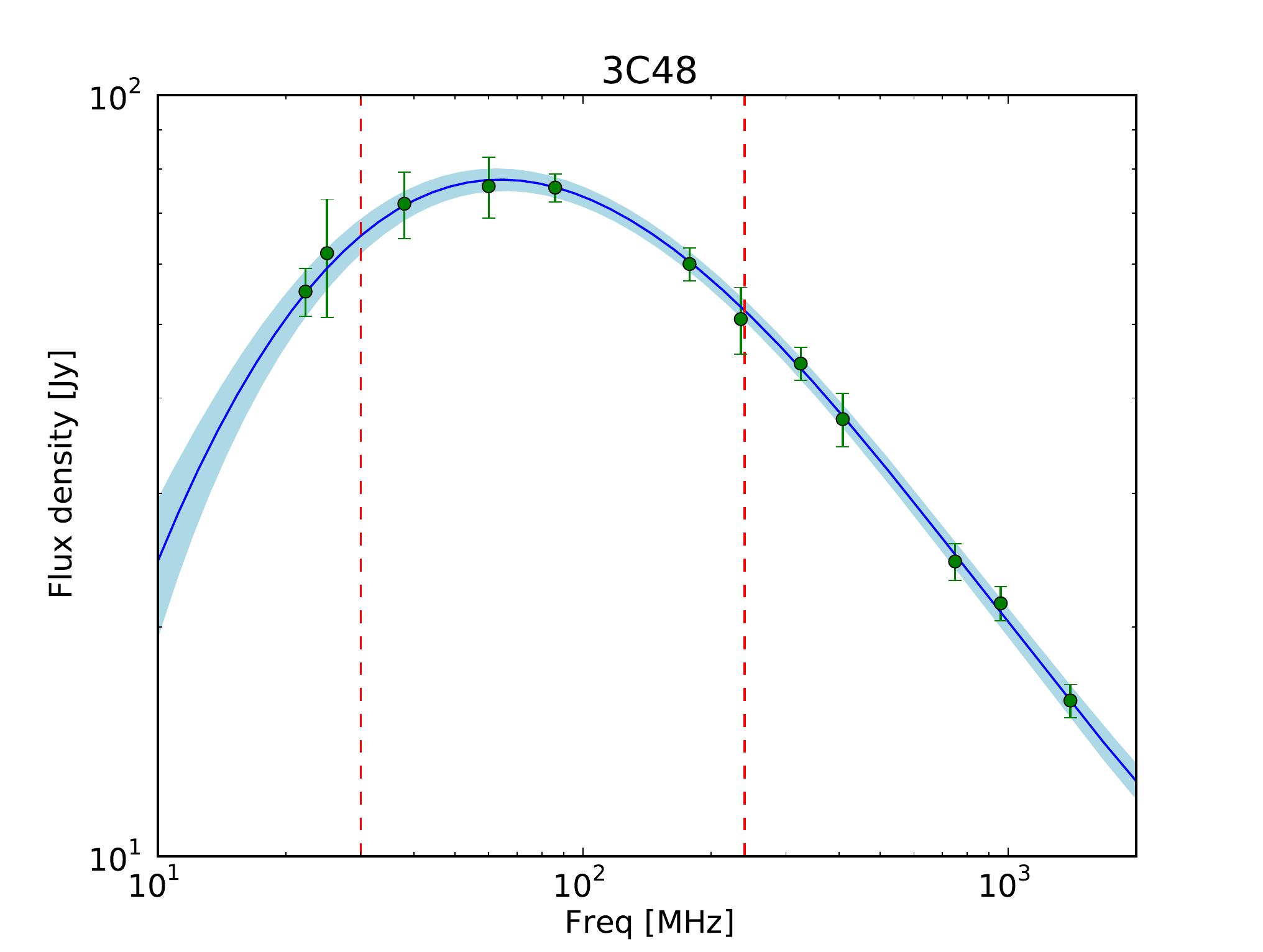}\includegraphics[width=0.33\textwidth]{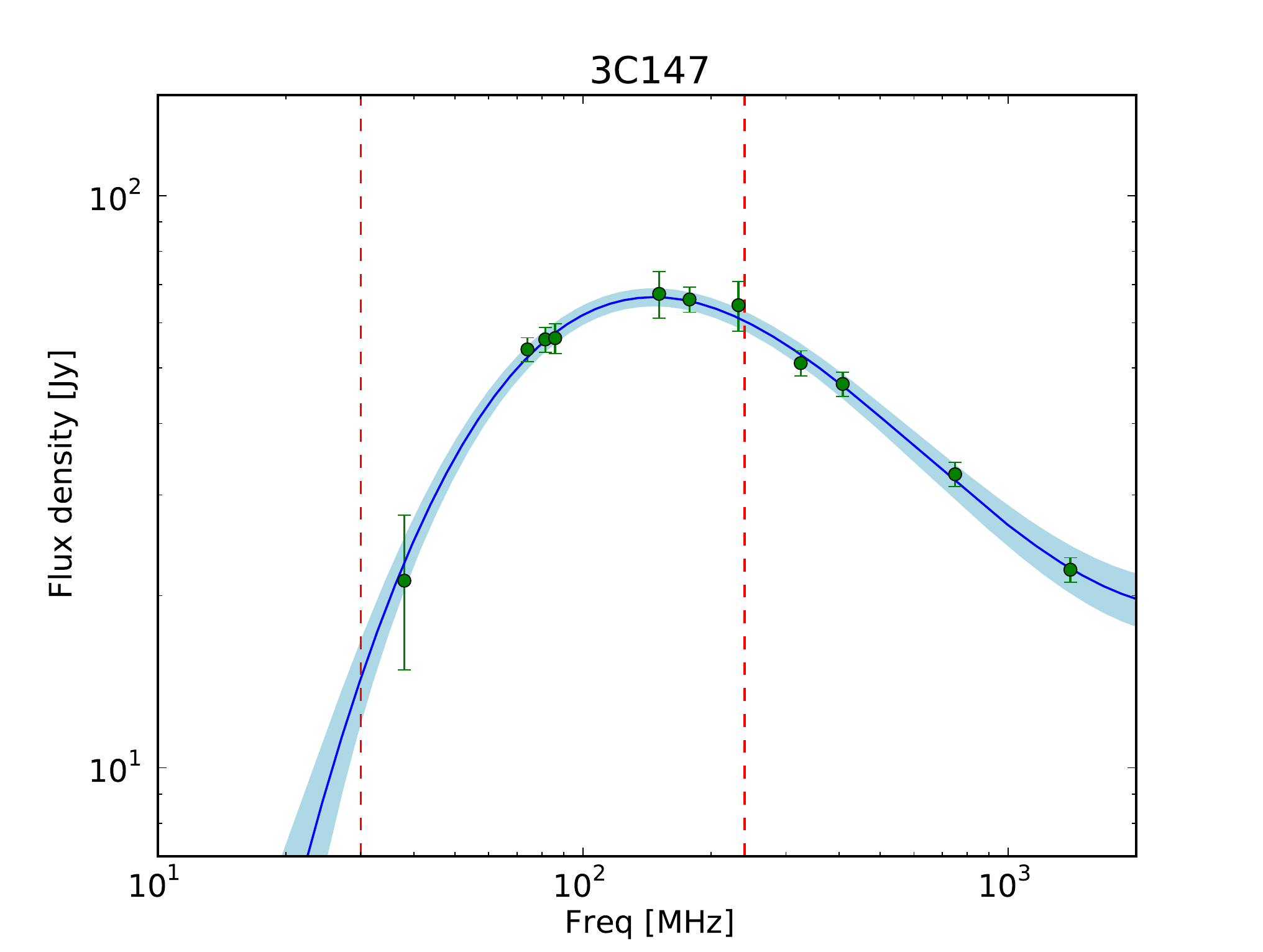}\includegraphics[width=0.33\textwidth]{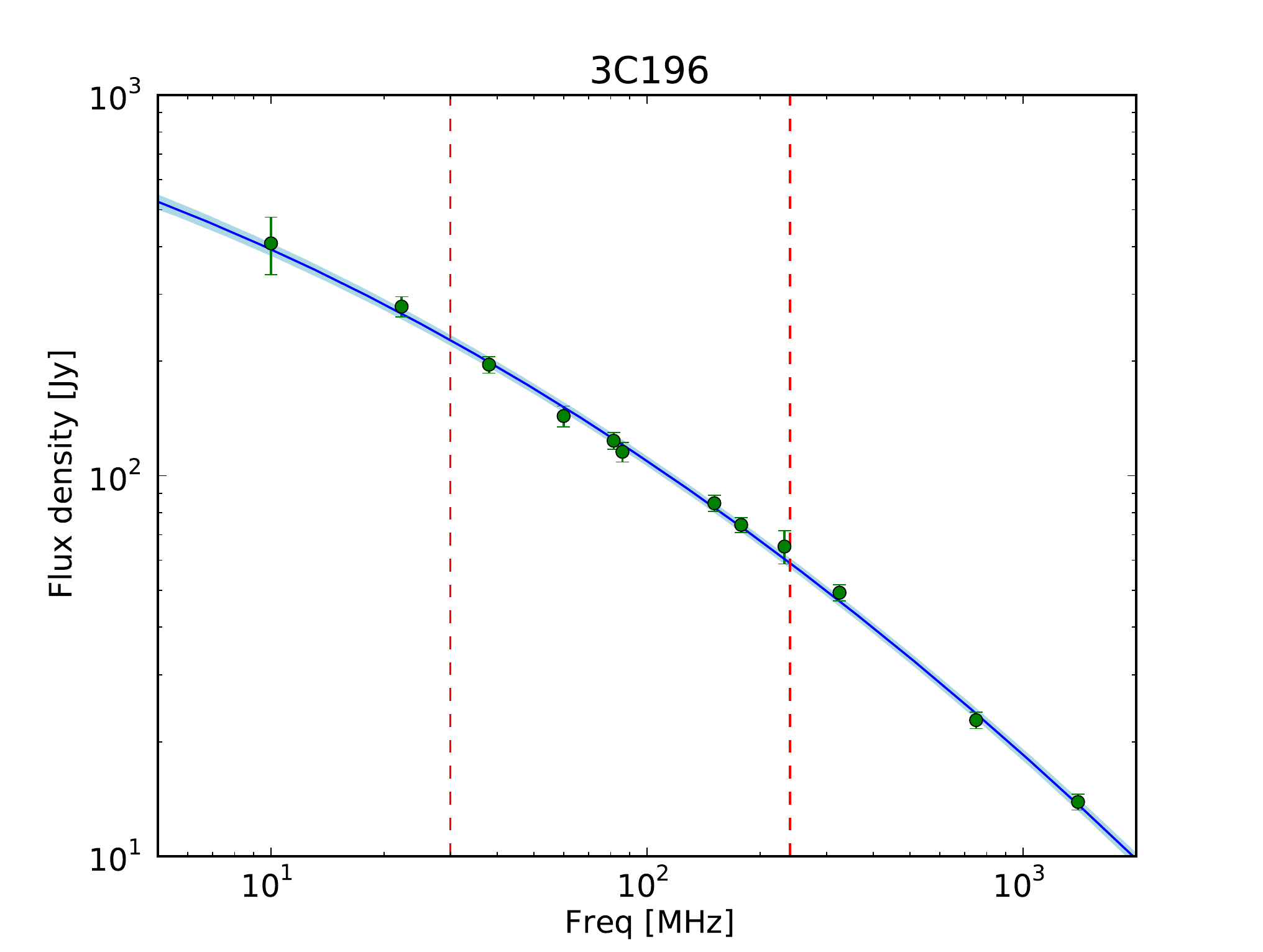}}
\centerline{\includegraphics[width=0.33\textwidth]{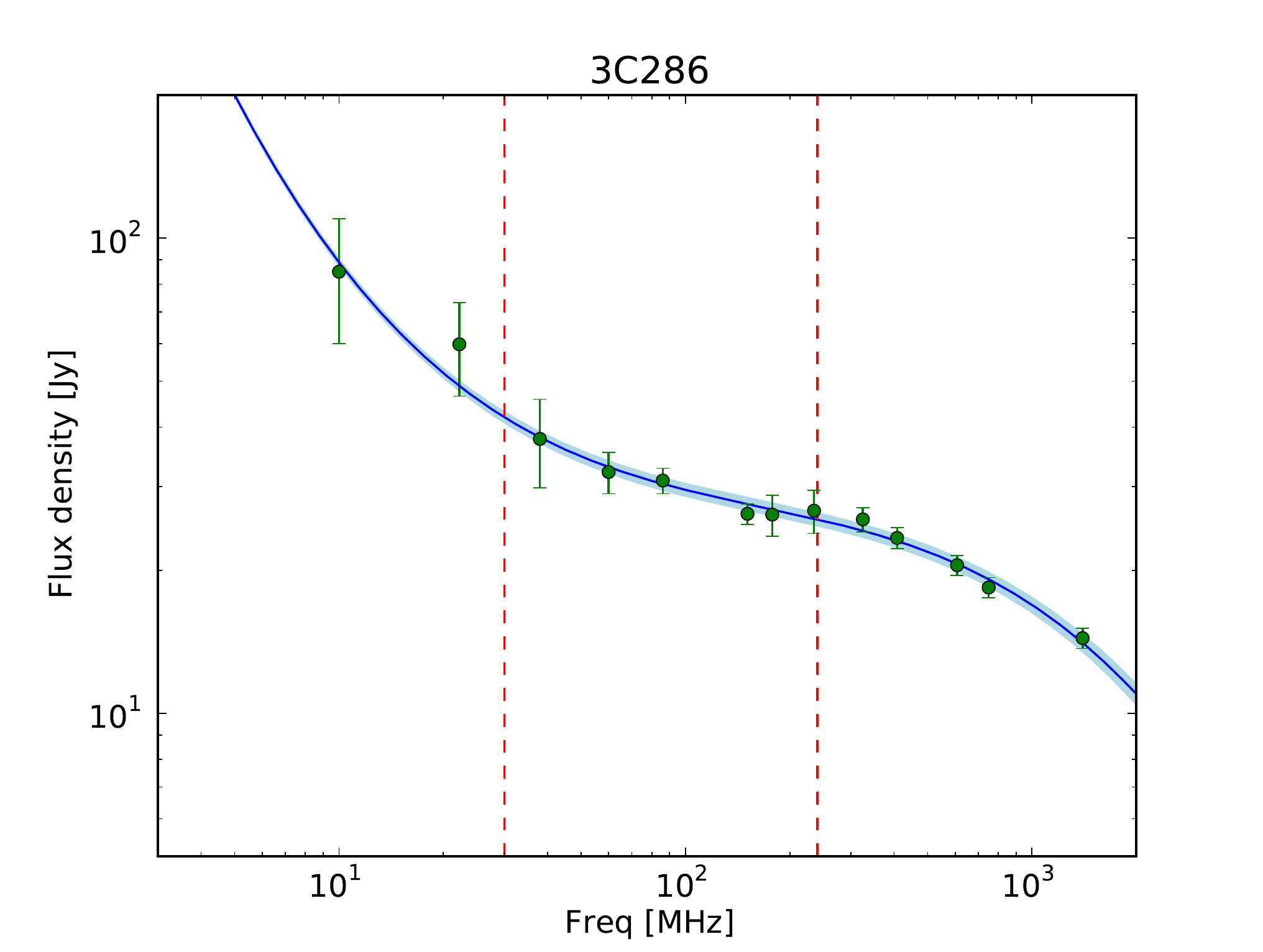}\includegraphics[width=0.33\textwidth]{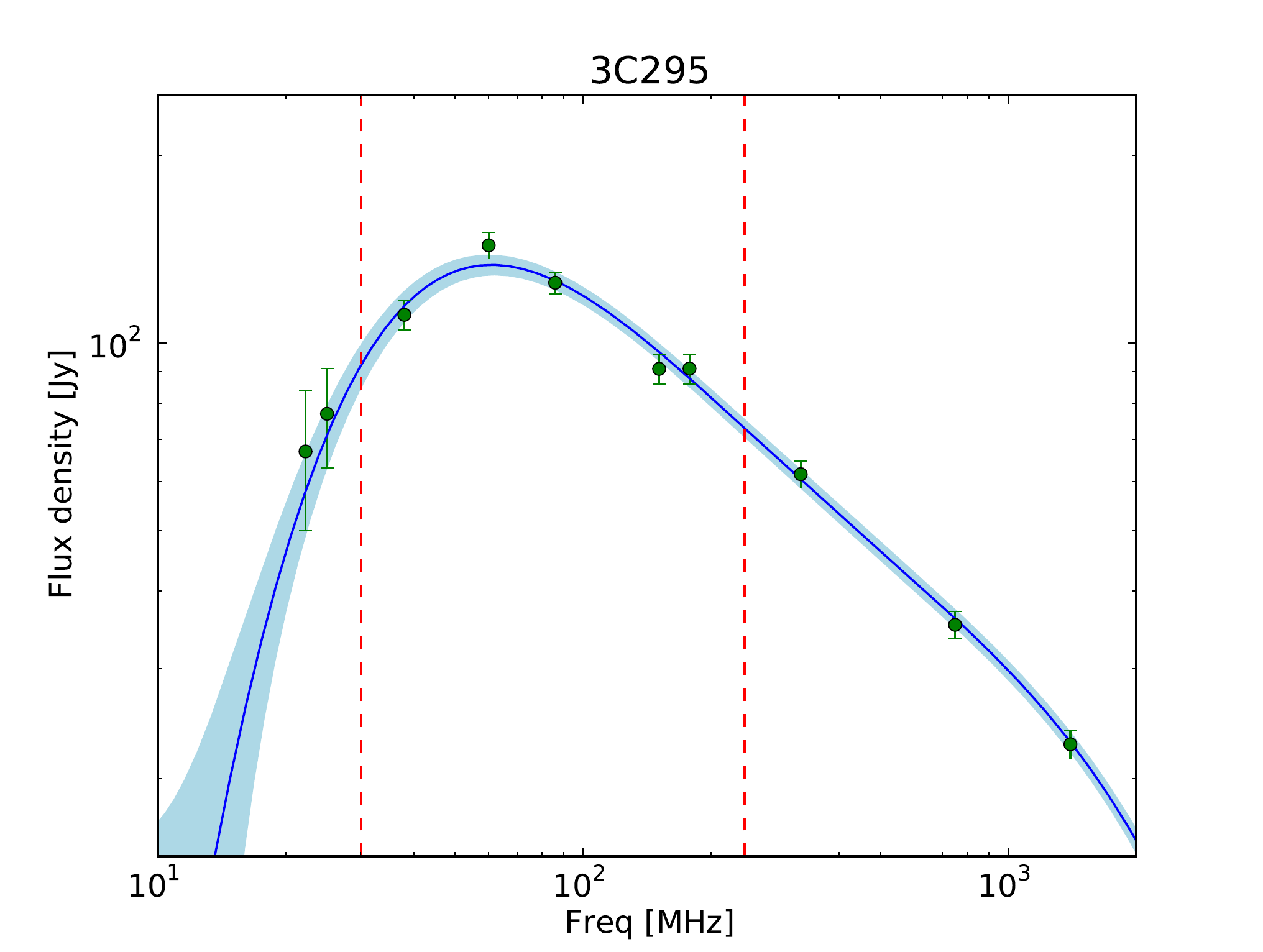}\includegraphics[width=0.33\textwidth]{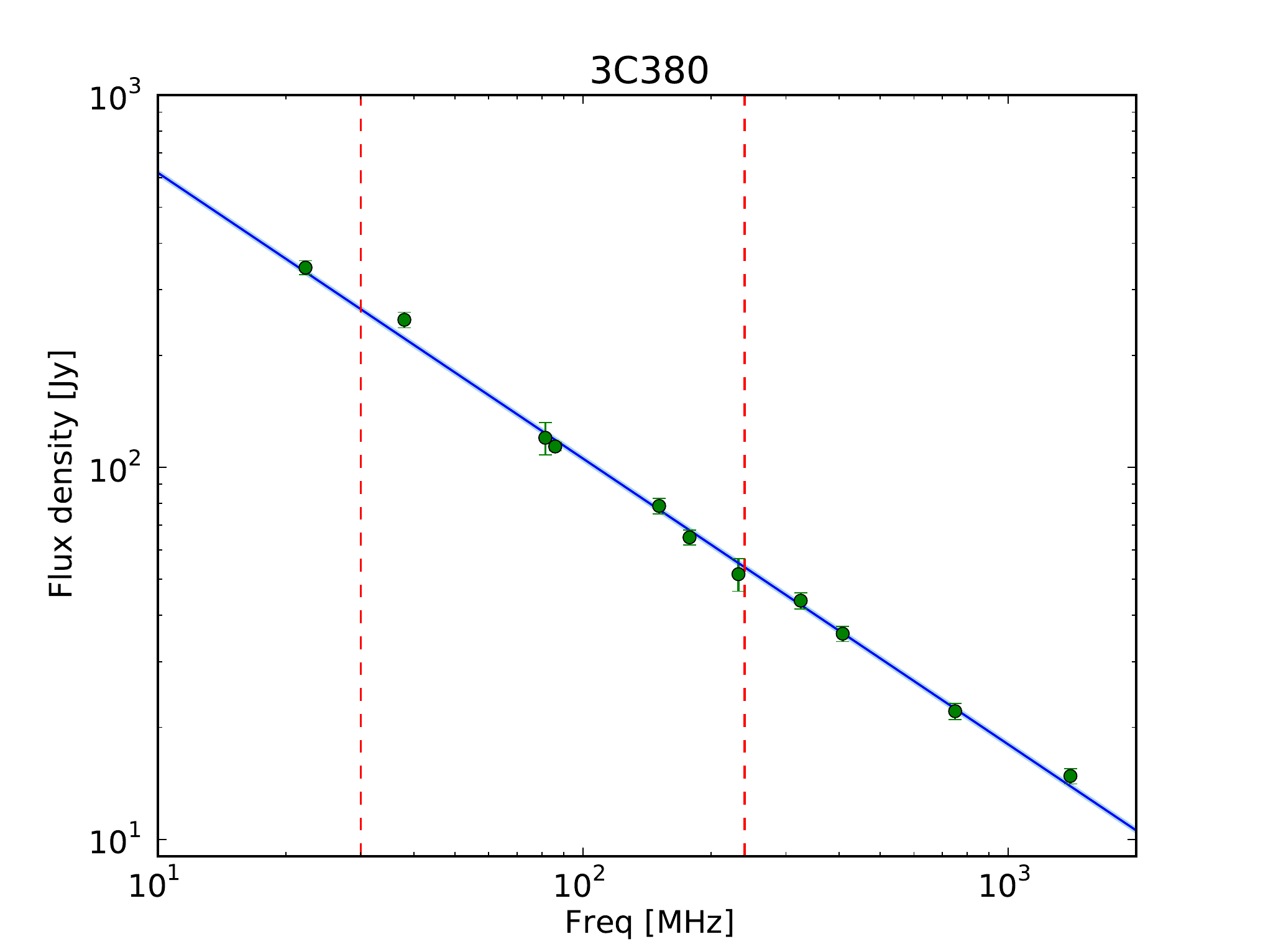}}
\caption{Best fitting models for calibrator sources. Data from the literature is shown in black with the selected best-fit model overlaid as a solid line. The area enclosed by the shaded region indicates the flux densities allowed at each frequency by one sigma uncertainties on the parameters. The edges of the LOFAR bandpass (low- and high-bands) are indicated by vertical dashed lines. \label{fig:best}}
\end{figure*}
\noindent
in order to retain Gaussian noise characteristics. Both determination of the optimal order ($N$) of polynomial model and maximum likelihood parameter estimation were performed using a Markov Chain Monte Carlo (MCMC) implementation. We used a simulated annealing method, through the {\sc metro} algorithm (Hobson \& Baldwin 2004), to employ a Bayesian inference approach, where Bayes' formula,
\begin{equation}
\nonumber {\rm Pr}(\Theta | D, H) \equiv \frac{ {\rm Pr}(D | \Theta,H) {\rm Pr}(\Theta | H)}{{\rm Pr}(D|H)}
\end{equation}
is used to test an hypothesis, $H$, parameterized by $\Theta$ using a set of data, $D$. Here ${\rm Pr}(\Theta | D, H) \equiv P(\Theta)$ is the posterior probability distribution of the parameters, ${\rm Pr}(D | \Theta,H) \equiv \mathcal{L}(\Theta)$ is the likelihood and ${\rm Pr}(\Theta | H) \equiv \Pi(\Theta)$ is the prior probability distribution, which in this case is simply used to restrict the volume of parameter space being sampled.

The Bayesian evidence, ${\rm Pr}(D|H) \equiv {Z}$, is a factor required for normalizing the posterior over the prior volume, such that
\begin{equation}
\nonumber {Z} = \int{ \mathcal{L}(\Theta) \Pi (\Theta) {\rm d}^{M}\Theta},
\end{equation} 
where $M$ is the dimensionality of the prior volume, here $M=N+1$. For parameter estimation the evidence factor can be neglected as it is independent of the model parameters. Maximum likelihood (ML) or maximum a posteriori (MAP) parameter values can be obtained by sampling the normalized distribution in each case to determine the peak in parameter space. However, in model selection the evidence becomes important for ranking different models based on a common dataset. It can be seen from the previous equation that the evidence represents the average of the likelihood over the prior, and therefore favors models with high likelihood values throughout the parameter space and penalizes models with regions of very low likelihood. This is equivalent to numerically implementing Occam's razor, whereby larger evidence values are returned for simple models (i.e. fewer parameters) with compact parameter spaces, compared to more complex models - unless the more complex model provides a significantly better fit to the data. 

Selecting between models, say $H_0$ and $H_1$, based on their evidence can be done using the ratio,
\begin{equation}
\nonumber \frac{ {\rm Pr}(H_0| D)}{{\rm Pr}(H_1|D)} = \frac{ {\rm Pr}(D | H_0) {\rm Pr}(H_0)}{ {\rm Pr}(D | H_1) {\rm Pr}(H_1)} = \frac{Z_0}{Z_1} \frac{{\rm Pr}(H_0)}{{\rm Pr}(H_1)}, 
\end{equation}
where ${{\rm Pr}(H_0)}/{{\rm Pr}(H_1)}$ is the ratio of prior probabilities. This ratio can be set before any conclusions have been drawn from the data; in many cases there is no reason to favor one particular model a priori and consequently this factor can be set to unity. In this circumstance the model selection can be based solely on the ratio of evidences. 

In this work, for each model, priors were assumed to be uniform and separable and ML (MAP) parameters were determined initially using the {\sc metro} sampling algorithm. Once parameter values had been determined, the evidence in each case, $Z$, was calculated over a $\pm3\sigma$ prior volume centered on the ML parameter values, with $\sigma$ determined for each parameter directly from the posterior distribution. The evidence calculation was repeated multiple times in each case in order to assess the variance of the evidence. Evidence ratios (also known as Bayes factors, or the odds) were then used to determine the optimal polynomial fit based on the Jeffreys scale (Jeffreys 1961), see \S~\ref{sec:selection}. In practice we take $\Delta\ln\,Z>1$ as our threshold for selecting the best model; this choice is justified in Section 5.

\subsection{Requirements for model selection}
\label{sec:selection}

The requirement to use a model of increased complexity (i.e. polynomial of higher order) depends upon the degree to which the evidence increases relative to the next lowest order, see Column [9] of Table~\ref{tab:res}. On the original Jeffreys scale (Jeffreys 1961) an increase of a factor of 3  (i.e. $\Delta  \ln Z \geq 3$) is considered substantial evidence to prefer the higher order model and can be considered equivalent to a 99.7\% confidence result. Revised versions of the Jeffreys scale (e.g. Gordon \& Trotta 2007) divide the level of support into categories where it is considered as either `inconclusive' ($\Delta \ln Z < 1$), `weak' ($1 \leq \Delta \ln Z \leq 2.5$), `moderate' ($2.5 \leq \Delta \ln Z \leq 5$), or `strong' ($\Delta \ln Z \geq 5$). 

\section{Results}

The ML parameters and evidence values for each polynomial fit to the source spectra are listed in Table~\ref{tab:res}. An example of the different orders is shown for 3C48 in Fig.~\ref{fig:3c48}. The poor fit of the linear  and $2^{\circ}$ polynomial model is evident by eye. This is also reflected in the values of the evidence for these models: an evidence ratio, and hence difference in the logarithm of the evidence, of $\ln Z_{3^{\circ}} - \ln Z_{1^{\circ}}>100$ indicates a definitive preference; a difference of $\ln Z_{3^{\circ}} - \ln Z_{2^{\circ}} = 3.16$ is substantial evidence for preferring the $3^{\circ}$ model above the $2^{\circ}$ model. The fractional evidence ratio, $\ln Z_{4^{\circ}} - \ln Z_{3^{\circ}} = -3.55$, between the $4^{\circ}$ and $3^{\circ}$ polynomial models indicates that the $3^{\circ}$ model is still preferred. In this case the goodness-of-fit is not diminished by the $4^{\circ}$ model, but there is no evidence in the data to support the use of the extra parameter and hence the model is penalized.

In general the results for this sample are easily interpreted, with Bayes factors of $\Delta \ln Z >3$ clearly indicating a preferred order of polynomial in most cases. When comparing different polynomial order fits to the 3C295 and 3C380 datasets the Bayes factors are less conclusive than in other cases. A difference of $\Delta \ln Z=1.6$ between the third and fourth order models in the case of 3C295 is intermediate to the `weak support' category. Although the support for moving to the higher order model is weak, it is not inconclusive and so in the context of the work here we choose to prefer the fourth order model. In the case of 3C380, a value of $\Delta \ln Z=0.64$ is securely in the inconclusive category and so we prefer the lower order model in this instance. Best fitting spectral models for the six calibrator sources are shown in Fig.~\ref{fig:best}. 
\begin{figure}
\centerline{\includegraphics[width=0.27\textwidth]{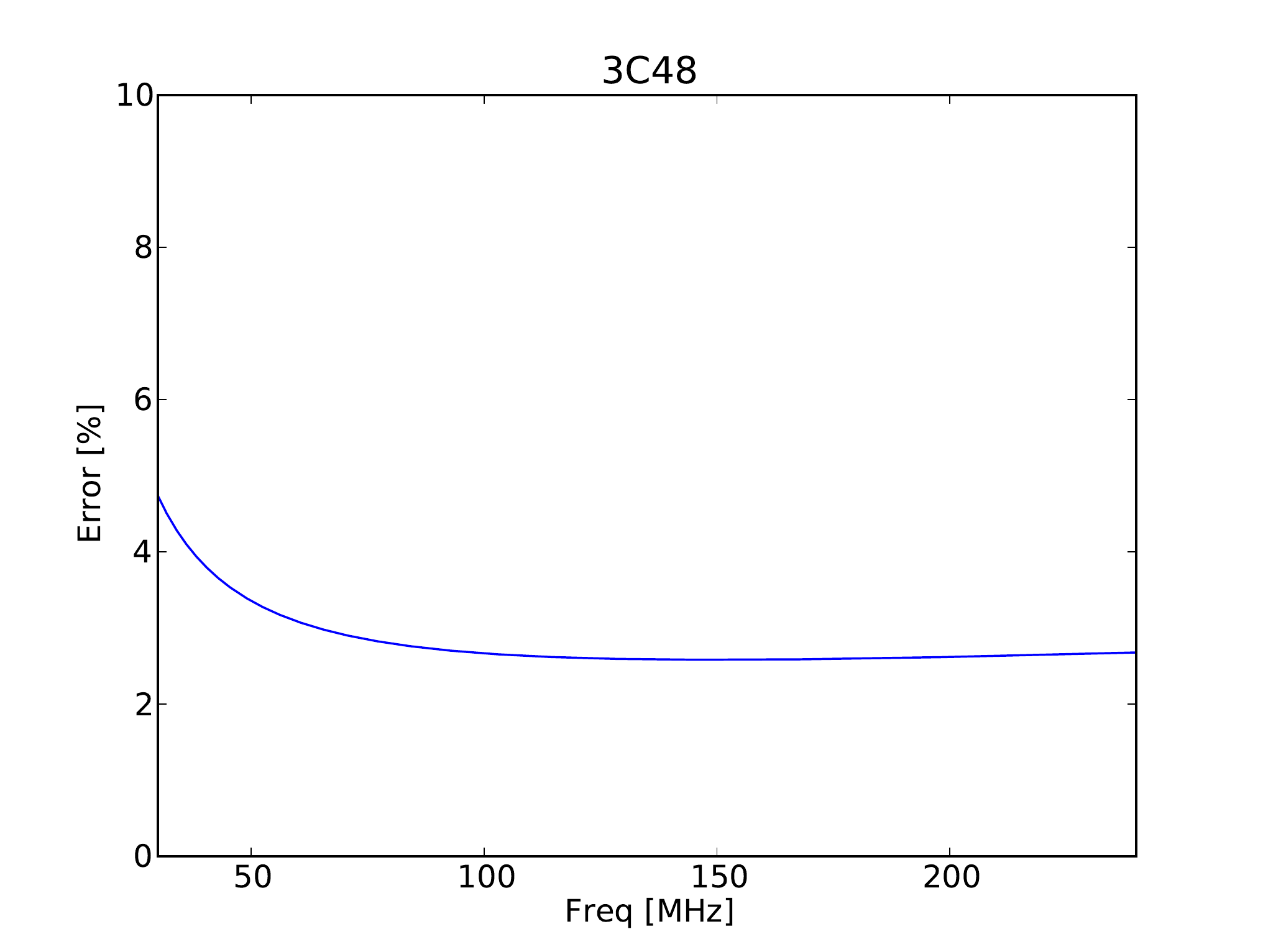}\includegraphics[width=0.27\textwidth]{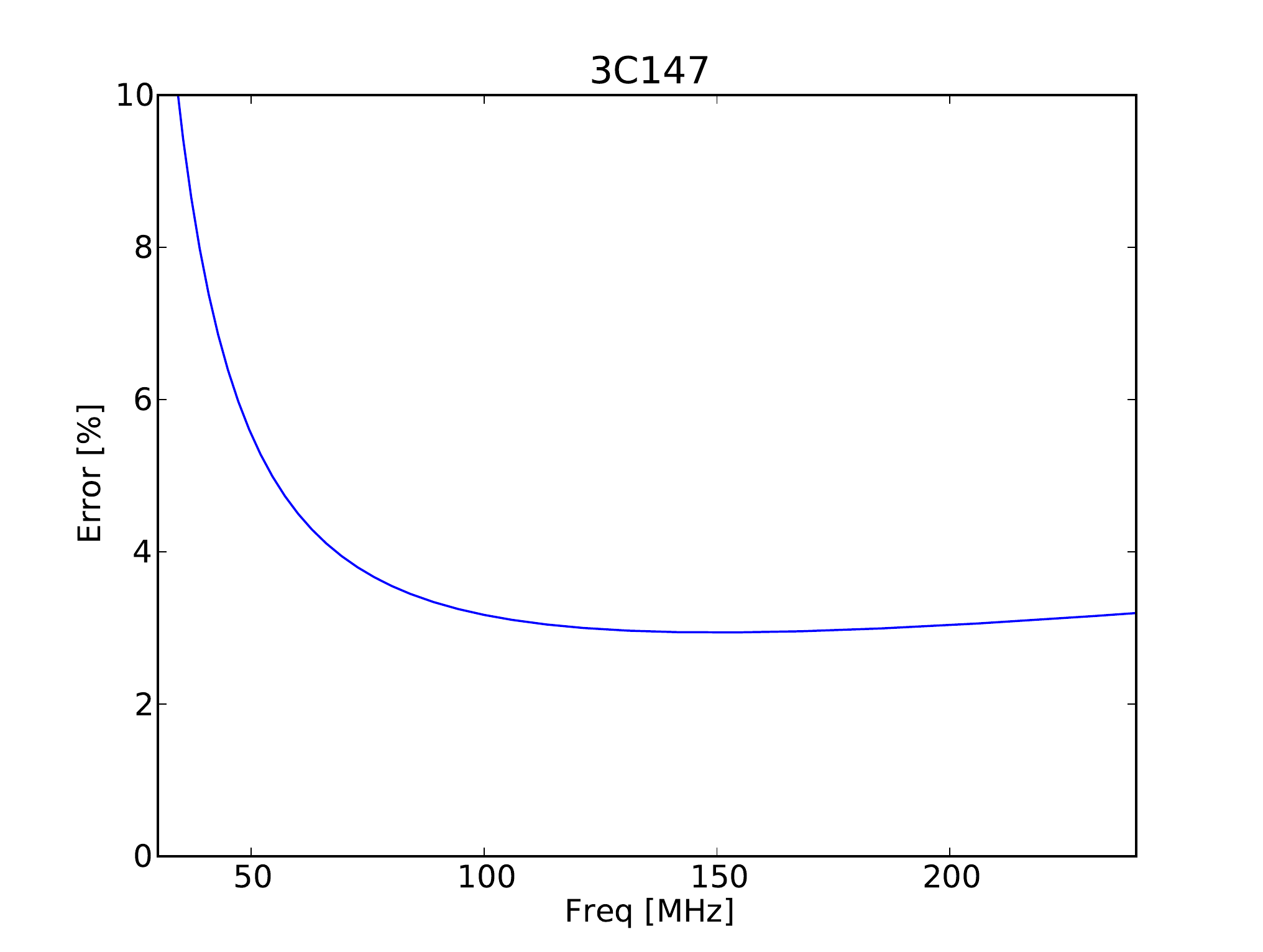}}
\centerline{\includegraphics[width=0.27\textwidth]{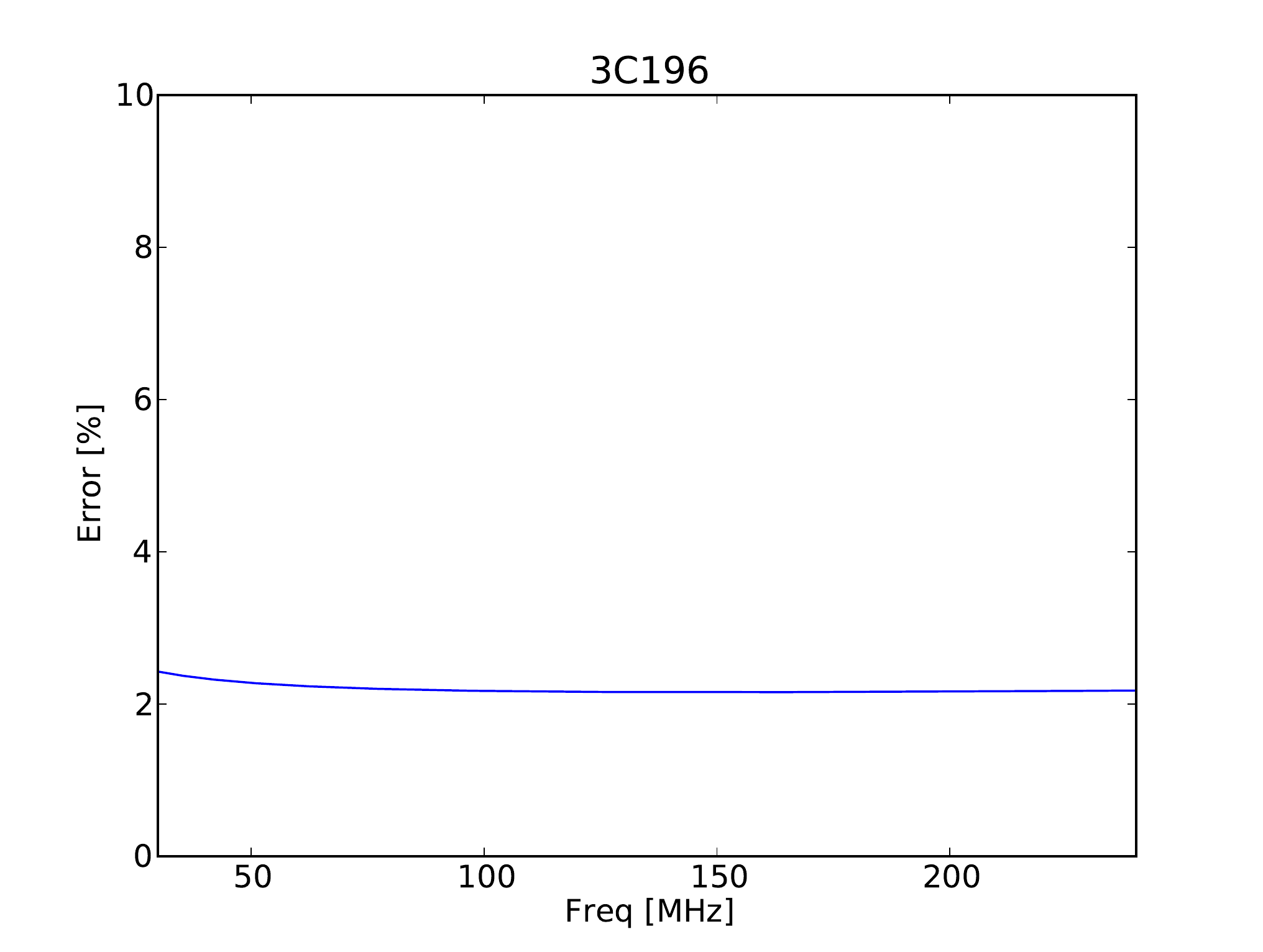}\includegraphics[width=0.27\textwidth]{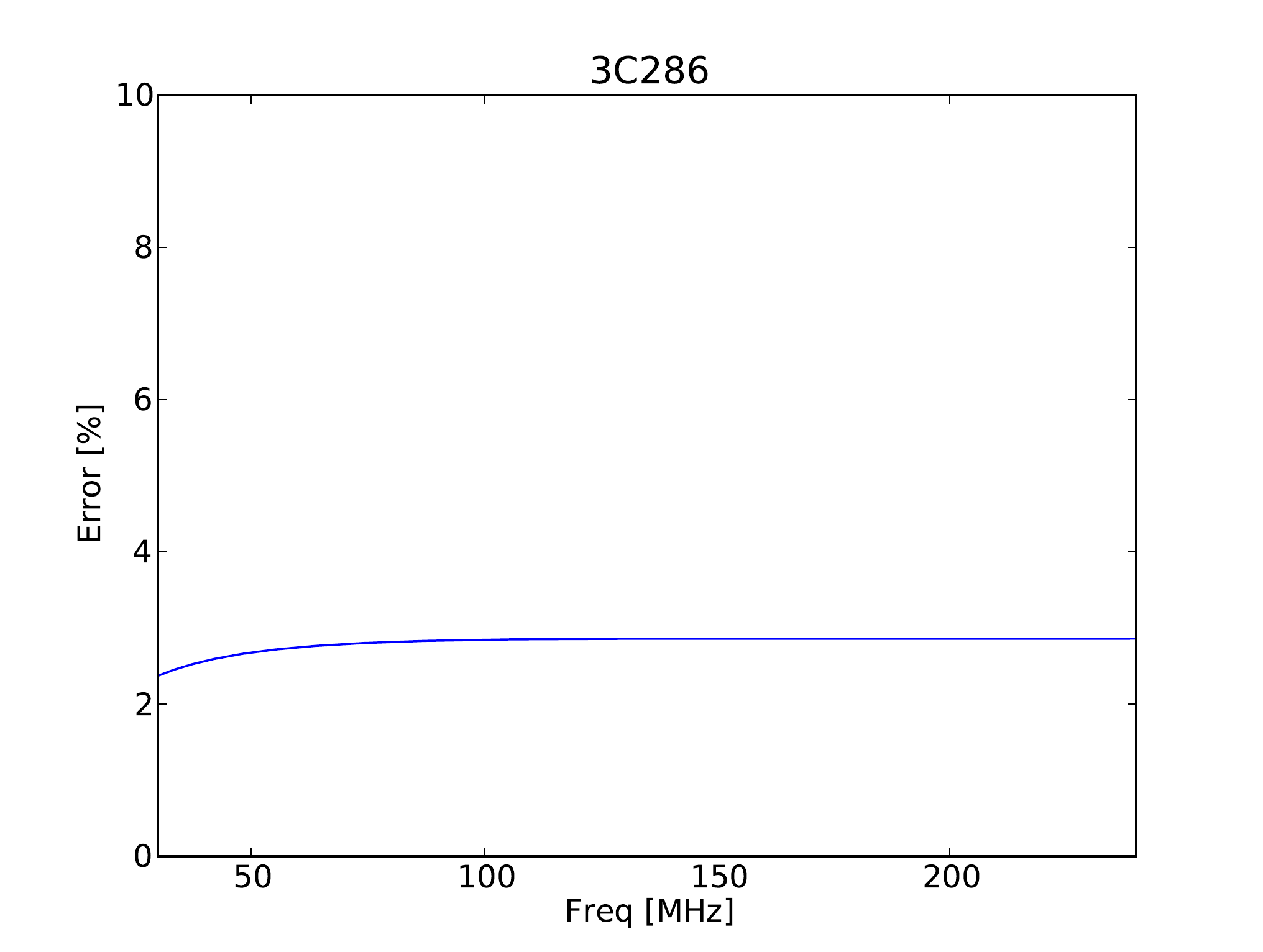}}
\centerline{\includegraphics[width=0.27\textwidth]{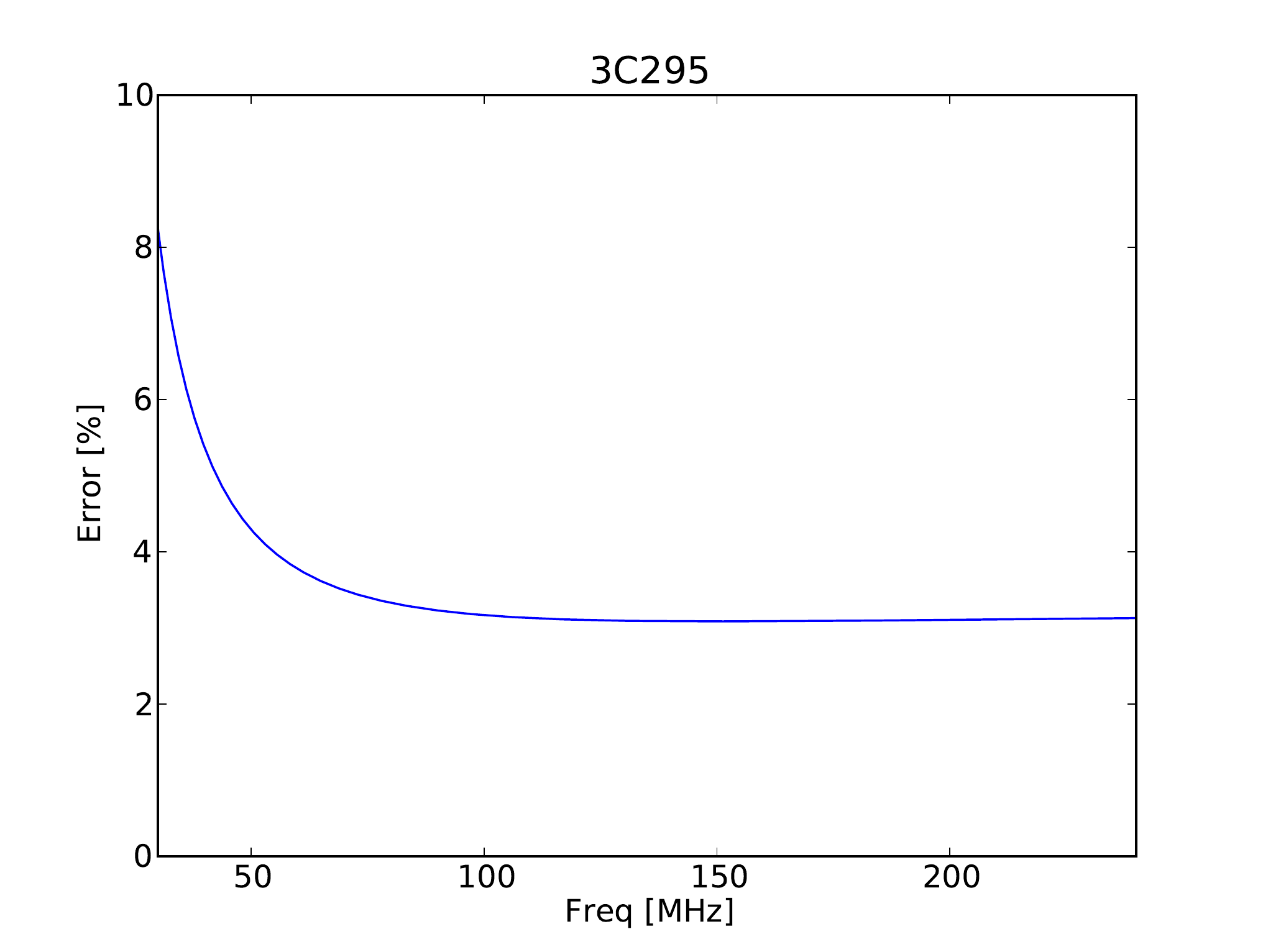}\includegraphics[width=0.27\textwidth]{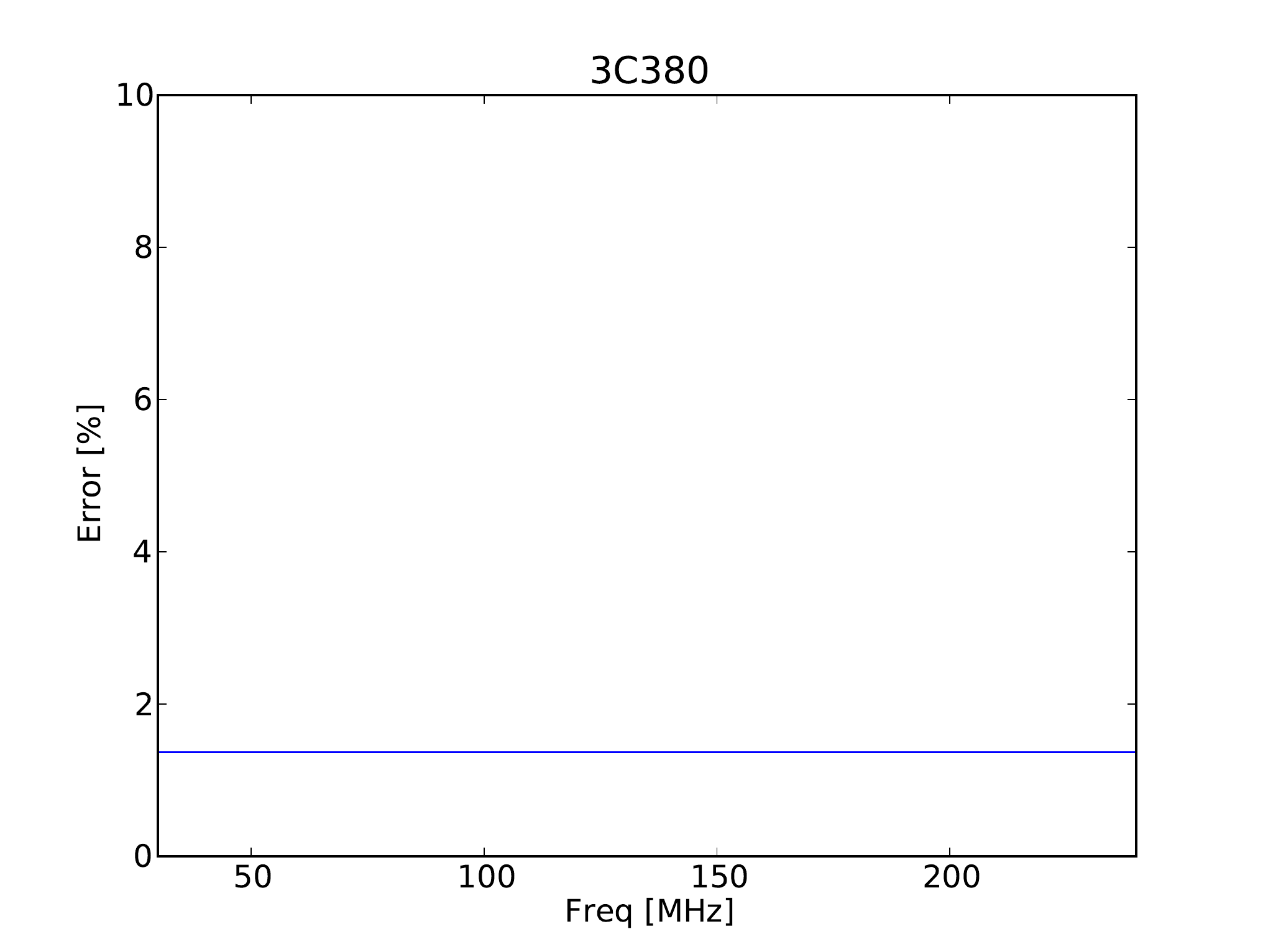}}
\caption{Percentage error in each model as a function of frequency from $30-240$\,MHz. \label{fig:errors}}
\end{figure}


\subsection{Error budget}
\label{sec:errors}

Errors on individual parameters for each fit were determined directly from the posterior distribution and are listed in Table~\ref{tab:res}. The uncertainty in the model due to these errors was derived analytically using differential error propagation and the $1\,\sigma$ bound on the model in each case is illustrated in Fig.~\ref{fig:best} as a blue shaded area. We illustrate the percentage error of each model as a function of frequency from $30-240$\,MHz (the LOFAR band) in Fig.~\ref{fig:errors}. It can be seen that from currently available data not all of the six calibrators are suitable for calibration at the low end of this frequency range (e.g. 3C147 \& 3C295), where their models possess high percentage uncertainty.
\begin{table*}
\footnotesize
\caption{Column [1] lists the order of the polynomial fit; columns [$2-6$] list the fitted ML polynomial coefficients; column [7] lists the reduced $\chi^2$ value; column [8] lists the natural logarithm of the evidence for a $3\sigma$ prior volume. The selected best-fit model is high-lighted in each case. \label{tab:res}}
\begin{tabular}{ccccccccr}
\hline\hline
Order & $A_0$ & $A_1$ & $A_2$ & $A_3$ & $A_4$ & $\chi^2_{\rm red}$ & $\ln Z$ & $\Delta(\ln Z)$ \\
\hline
\textbf{3C48}: &&&&&&&\\
1$^{\circ}$ & $43.874\pm0.879$ & $-0.349\pm0.011$ & $-$ & $-$ & $-$ & 21.68 & $-135.00\pm0.03$ & $-$\\
2$^{\circ}$ & $62.821\pm1.642$ & $-0.284\pm0.015$ & $-0.374\pm0.026$ & $-$ & $-$ & 1.16 & $-33.02\pm0.06$ & 101.98\\
\rowcolor{green} 3$^{\circ}$ & $64.768\pm1.761$ & $-0.387\pm0.039$ & $-0.420\pm0.031$ & $0.181\pm0.060$ & $-$ & 0.10 & $-29.86\pm0.23$ & 3.16\\
4$^{\circ}$ & $63.910\pm1.864$ & $-0.394\pm0.045$ & $-0.391\pm0.093$ & $0.185\pm0.075$ & $-0.014\pm0.118$ & 0.15 & $-33.41\pm0.47$ & $-3.55$\\
\textbf{3C147}: &&&&&&&\\ 
2$^{\circ}$ & $60.517\pm1.474$ & $0.016\pm0.028$ & $-0.514\pm0.046$ & $-$ & $-$ & 2.35 & $-35.28\pm0.14$ & $-$ \\
\rowcolor{green} 3$^{\circ}$ & $66.738\pm2.490$ & $-0.022\pm0.030$ & $-1.012\pm0.167$ & $0.549\pm0.170$ & $-$ & 0.24 & $-29.59\pm0.47$ & 5.69\\
4$^{\circ}$ & $66.494\pm1.915$ & $-0.041\pm0.046$ & $-0.952\pm0.109$ & $0.625\pm0.245$ & $-0.124\pm0.249$ & 0.26 & $-30.15\pm0.48$ & $-0.59$ \\
\textbf{3C196}: &&&&&&&\\
1$^{\circ}$ & $76.641\pm1.227$ & $-0.719\pm0.012$ & $-$ & $-$ & $-$ & 2.80 & $-47.25\pm0.08$ & $-$ \\
\rowcolor{green} 2$^{\circ}$ & $83.084\pm1.862$ & $-0.699\pm0.014$ & $-0.110\pm0.024$ & $-$ & $-$ & 0.51 & $-36.89\pm0.09$ & 10.36\\
3$^{\circ}$ & $83.011\pm1.787$ & $-0.676\pm0.029$ & $-0.107\pm0.023$ & $-0.039\pm0.041$ & $-$ & 0.50 & $-38.16\pm0.14$ & $-1.27$\\
4$^{\circ}$ & $83.776\pm2.214$ & $-0.677\pm0.033$ & $-0.139\pm0.073$ & $-0.027\pm0.045$ & $0.035\pm0.073$ & 0.54 & $-40.37\pm0.50$ & $-2.21$\\
\textbf{3C286}: &&&&&&&\\ 
1$^{\circ}$ & $27.893\pm0.653$ & $-0.258\pm0.017$ & $-$ & $-$ & $-$ & 1.46 & $-33.93\pm0.08$ & $-$ \\
2$^{\circ}$ & $28.230\pm0.708$ & $-0.208\pm0.035$ & $-0.077\pm0.045$ & $-$ & $-$ & 1.32 & $-34.09\pm0.14$ & $-0.16$\\
\rowcolor{green} 3$^{\circ}$ & $27.477\pm0.746$ & $-0.158\pm0.033$ & $0.032\pm0.043$ & $-0.180\pm0.052$ & $-$ & 0.42 & $-30.51\pm0.21$ & 3.58\\
4$^{\circ}$ & $27.591\pm0.911$ & $-0.144\pm0.038$ & $0.005\pm0.097$ & $-0.187\pm0.054$ & $0.021\pm0.086$ & 0.48 & $-32.58\pm0.49$ & $-2.07$\\
\textbf{3C295}: &&&&&&&\\
2$^{\circ}$ & $97.489\pm2.177$ & $-0.347\pm0.016$ & $-0.362\pm0.028$ & $-$ & $-$ & 6.12 & $-49.27\pm0.07$ & $-$\\
3$^{\circ}$ & $100.950\pm2.454$ & $-0.517\pm0.035$ & $-0.497\pm0.041$ & $0.360\pm0.066$ & $-$ & 1.85 & $-35.46\pm0.26$ & $13.81$ \\
\rowcolor{green} 4$^{\circ}$ & $97.763\pm2.787$ & $-0.582\pm0.045$ & $-0.298\pm0.085$ & $0.583\pm0.116$ & $-0.363\pm0.137$ & 1.00 & $-33.86\pm0.34$ & $1.60$\\
5$^{\circ}$ & $-$ & $-$ & $-$ & $-$ & $-$ & 1.30 & $-36.33\pm0.22$ & $-2.47$\\
\textbf{3C380}: &&&&&&&\\
\rowcolor{green} 1$^{\circ}$ & $77.352\pm1.164$ & $-0.767\pm0.013$ & $-$ & $-$ & $-$ & 1.20 & $-38.11\pm0.06$ & $-$\\
2$^{\circ}$ & $75.682\pm1.537$ & $-0.772\pm0.012$ & $0.039\pm0.021$ & $-$ & $-$ & 1.02 & $-37.47\pm0.10$ & $0.64$\\
3$^{\circ}$ & $75.233\pm1.483$ & $-0.788\pm0.033$ & $0.041\pm0.020$ & $0.024\pm0.047$ & $-$ & 1.11 & $-39.22\pm0.15$ & $-1.75$\\
4$^{\circ}$ & $74.386\pm1.595$ & $-0.787\pm0.034$ & $0.104\pm0.067$ & $0.030\pm0.051$ & $-0.084\pm0.082$ & 1.14 & $-40.96\pm0.40$ & $-1.74$\\

\hline
\end{tabular}
\end{table*}

\subsection{Notes on Individual Sources}

\noindent
\textbf{3C380} This source has data at 10\,MHz in Bridle \& Purton (1968) but the very low flux density (168\,Jy) indicates that the spectrum turns over sharply below 20\,MHz. The effect of this turn over is marginal above 30\,, but would require significantly increased complexity in the model. Consequently these data have been excluded from the fit.

\section{Conclusions}

We have presented parameterized broadband spectral models for six bright radio sources selected from the 3C survey between 30-300\,MHz, spread in Right Ascension from $0-24$\,hours. For each source, data from the literature have been compiled and tied to a common flux density scale. These data have then been used to parameterize an analytic polynomial spectral calibration model. The best fitting polynomial model order in each case has been determined using the ratio of the Bayesian evidence for the candidate models. Maximum likelihood parameter values with associated errors have been presented. The percentage error in each model as a function of frequency has been derived and is illustrated in \S~\ref{sec:errors}. These spectral models are intended as an initial reference for science quality data from the new generation of low frequency telescopes, such as LOFAR, now coming on line. In this context we have shown that two of these sources lead to unacceptably high flux scale uncertainty at frequencies below 70\,MHz (3C147 \& 3C295), and we also note that 3C380 may be unsuitable for precision calibration at higher frequencies where its angular extent becomes an issue.

\section{ACKNOWLEDGEMENTS}
We thank Julia Riley and Ger de~Bruyn for useful discussions and Rick Perley for his careful review of the manuscript. This research has made use of the Astrophysical CATalogs Support System (CATS) and the NASA/IPAC Extragalactic Database (NED).

\end{document}